\documentclass[a4paper]{report}
\usepackage[utf8]{inputenc}
\usepackage[T1]{fontenc}
\usepackage{RJournal}
\usepackage{amsmath,amssymb,array}
\usepackage{booktabs}

\usepackage{longtable}

\newlength{\cslhangindent}
\newlength{\csllabelwidth}
\newlength{\cslentryspacingunit} 
  {}%
  {\par}
 {
  \setlength{\parindent}{0pt}
  \ifodd #1
  \let\oldpar\par
  \def\par{\hangindent=\cslhangindent\oldpar}
  \fi
  \setlength{\parskip}{#2\cslentryspacingunit}
 }%
 {}
\usepackage{calc}

\begin{document}

\sectionhead{Contributed research article}
\volume{XX}
\volnumber{YY}
\year{20ZZ}
\month{AAAA}

\begin{article}
\title{midr: Learning from Black-Box Models by Maximum Interpretation Decomposition}

\author{by Ryoichi Asashiba, Reiji Kozuma, and Hirokazu Iwasawa}

\maketitle

\abstract{%
\quad The use of appropriate methods of Interpretable Machine Learning (IML) and eXplainable Artificial Intelligence (XAI) is essential for adopting black-box predictive models in fields where model and prediction explainability is required.
As a novel tool for interpreting black-box models, we introduce the R package \pkg{midr}, which implements Maximum Interpretation Decomposition (MID). MID is a functional decomposition approach that derives a low-order additive representation of a black-box model by minimizing the squared error between the model's prediction function and this additive representation.
\pkg{midr} enables learning from black-box models by constructing a global surrogate model with advanced analytical capabilities. After reviewing related work and the theoretical foundation of MID, we demonstrate the package's usage and discuss some of its key features.
}

\hypertarget{introduction}{%
\section{Introduction}\label{introduction}}

A predictive model is characterized by a prediction function that maps observations \(\mathbf{x}=\left(x_1,\dots,x_d\right)\) to predictions \(f(\mathbf{x})\). When it is difficult to understand the behavior of the prediction function, i.e., the correspondence between \(\mathbf{x}\) and \(f(\mathbf{x})\), the model is said to be a black-box. To understand and explain the behavior of such black-box models, various interpretation methods have been proposed \citep{molnar2025, biecek2021}, collectively referred to as Interpretable Machine Learning (IML) or eXplainable Artificial Intelligence (XAI).

Interpretation methods can be classified into \emph{model-specific} or \emph{model-agnostic} approaches. Model-specific methods use the computational processes and outputs unique to the fitting algorithm of a predictive model. In contrast, model-agnostic methods analyze a black-box model based solely on its input-output behavior, independent of its specific algorithm or internal structure, often by observing how predictions change in response to modifications of the input data. While model-agnostic methods typically demand more computational time and resources than model-specific methods, they have the notable advantage of enabling comparisons of interpretation results across different types of predictive models.

From another perspective, the interpretations themselves can be classified by their scope as either \emph{local} or \emph{global}. Local interpretations focus on individual predictions, whereas global interpretations address the average behavior of the prediction function over the empirical or probability distribution of features. Some interpretation methods can be applied to both local and global contexts. For instance, ICE \citep{goldstein2015} can be used for local interpretations, such as examining the sensitivity of a specific prediction, as well as for global interpretations, such as detecting the presence of interactions between features. Similarly, SHAP \citep{lundberg2017} can be used for local interpretations, such as decomposing a specific prediction, and for global interpretations, such as summarizing the average importance or effects of features.

This paper introduces the R package \pkg{midr} \citep{midr}, which implements an interpretation method for predictive models based on Maximum Interpretation Decomposition (MID) \citep{iwasawa2025}. MID is a functional decomposition technique that generates an additive approximation of a prediction function by minimizing the squared error between them, subject to certain constraints. It also serves as a model-agnostic interpretation method applicable to both global and local scopes of interpretation. In essence, \pkg{midr} provides functionalities to create global surrogate models using first- or second-order MID of a given predictive model and to summarize and visualize these surrogate models from multiple perspectives.

The paper proceeds as follows: Section \protect\hyperlink{related-work}{2} provides a brief overview of various interpretation methods, focusing on visualization techniques and their implementations. Section \protect\hyperlink{maximum-interpretation-decomposition}{3} outlines the definition, numerical estimation, and visualization strategies of MID. Section \protect\hyperlink{constructing-mid-in-r}{4} demonstrates the use of the package for analyzing black-box models. Section \protect\hyperlink{discussion}{5} discusses two key features of MID as an interpretation method, supported by numerical simulations. Finally, \protect\hyperlink{appendix-numerical-estimation-of-mid}{Appendix} provides a detailed explanation of the estimation method of MID implemented in the \pkg{midr} package.

\hypertarget{related-work}{%
\section{Related work}\label{related-work}}

Many R packages and Python libraries provide a variety of functionalities for interpreting predictive models \citep{maksymiuk2020}. Among the R packages available on CRAN, \CRANpkg{iml} \citep{iml}, \CRANpkg{DALEX} \citep{DALEX}, and \CRANpkg{flashlight} \citep{flashlight} are particularly well known as versatile packages implementing numerous interpretation methods. In practical applications, SHAP is the most commonly used method \citep{saarela2024} and has several implementations, including the Python library \pkg{shap} \citep{lundberg2017}.

The following subsections provide an overview of widely adopted interpretation methods for predictive models, as well as their implementations in R and Python.

\hypertarget{functional-decomposition}{%
\subsection*{Functional decomposition}\label{functional-decomposition}}
\addcontentsline{toc}{subsection}{Functional decomposition}

Several interpretation methods have been proposed to approximate the complex prediction function \(f\) of black-box models as a sum of functions with fewer variables, as shown in \eqref{eq:Decomposition}.
\begin{equation}
f(\mathbf{x}) \approx f_{\emptyset} + \sum\limits_{j} f_j(x_j) + \sum\limits_{j<k} f_{jk}(x_j, x_k) + \sum\limits_{j<k<l} f_{jkl}(x_j, x_k, x_l) + \cdots
  \label{eq:Decomposition}
\end{equation}
Here, \(f_{\emptyset}\) is the intercept, \(f_j(x_j)\) is the main effect of feature \(j\), and \(f_{jk}(x_j, x_k)\) is the second-order interaction effect of the feature pair \((j, k)\). The third- or higher-order terms are defined analogously. This approach to model interpretation is known as functional decomposition.

The main effects can be visualized by line plots or bar plots based on
\begin{equation}
\left\{ \left(x_j,\ f_j(x_j)\right)\;\middle|\;x_j^{\min} \leq x_j \leq x_j^{\max} \right\}
  \label{eq:MainEffectPlot}
\end{equation}
where \(x_j^{\min}\) and \(x_j^{\max}\) represent the minimum and maximum observed values of feature \(j\), and the second-order interactions are visualized by surface plots, contour plots, or heatmaps based on
\begin{equation}
\left\{ \left( x_j,\ x_k,\ f_{jk}(x_j,x_k) \right)\;\middle|\;x_j^{\min} \leq x_j \leq x_j^{\max},
\ x_k^{\min} \leq x_k \leq x_k^{\max} \right\}
  \label{eq:InteractionPlot}
\end{equation}

When the prediction function of a predictive model is additive, the main effects and second-order interactions can be directly visualized. R packages such as \CRANpkg{gam} \citep{gam} and \CRANpkg{mgcv} \citep{mgcv}, which implement Generalized Additive Models \citep{hastie1990}, facilitate the visualization of feature effects in fitted GAMs. For predictive models based on ensembles of decision trees, restricting the maximum depth of individual trees to \(h\) enables the prediction function, defined as the weighted sum of the trees, to be decomposed into a sum of main and interaction effects of order \(h\) or less. Explainable Boosting Machine (EBM) \citep{lou2013}, a predictive model that applies this concept, is implemented in the Python library \pkg{interpret} \citep{interpret}. These decomposition techniques are model-specific, and alternative approaches are needed to decompose general prediction functions.

The Partial Dependence Plot (PDP) is a model-agnostic interpretation method proposed by \citet{friedman2001}. The PDP for a set of features \(J\) is calculated as the expected value of a prediction function based on the marginal distribution of the other features \(\mathbf{X}_{\setminus J}\) at each point where observations \(\mathbf{x}_J\) can take values. Precisely, it is defined and estimated as follows:
\begin{align}
\mathrm{PD}_J(\mathbf{x}_J) &=
\mathbb{E} { \left[f(\mathbf{x}_J, \mathbf{X}_{\setminus J}) \right]}
  \label{eq:PD1} \\
\mathrm{\hat{PD}}_J(\mathbf{x}_J) &= \frac{1}{n} \sum_{i=1}^{n} f(\mathbf{x}_J, \mathbf{x}_{i,\setminus J})
  \label{eq:PD2}
\end{align}
However, when using PD as a functional decomposition method, it is necessary to center each effect and subtract lower order effects. Precisely, the main and interaction effects are defined as follows:
\begin{align}
f^{\text{pd}}_j(x_j) &= \mathrm{PD}_j(x_j) - \mathbb{E}{\left[\mathrm{PD}_j(X_j)\right]}
  \label{eq:PDBasedDecomposition1} \\
f^{\mathrm{pd}}_{jk}(x_j,x_k) &= \mathrm{PD}_{jk}(x_j,x_k) - \mathbb{E}{\left[\mathrm{PD}_{jk}(X_j,X_k)\right]} - f^{\mathrm{pd}}_j(x_j) - f^{\mathrm{pd}}_k(x_k)
  \label{eq:PDBasedDecomposition2}
\end{align}
PDP has been implemented in many R packages including \CRANpkg{pdp} \citep{pdp} as a method for visualizing feature effects.

One of the challenges of PDP is that, when features are correlated, its reliance on marginal distributions can lead to averaging predictions over unrealistic data instances, which may distort the estimated effects. This issue is discussed in detail in Section \protect\hyperlink{discussion}{5}. As an alternative to PDP, \citet{apley2020} proposed Accumulated Local Effect (ALE). For a differentiable prediction function \(f\), the uncentered ALE main effect of feature \(j\) is defined by integrating the expected partial derivative of \(f\) with respect to \(x_j\), where the expectation is conditional on \(X_j = z_j\):
\begin{equation}
\mathrm{ALE}_j(x_j) = \int_{x_j^{\min}}^{x_j} \mathbb{E} {\left[ \frac{\partial f}{\partial x_j} {(\mathbf{X})}\;\middle|\;X_j = z_j \right]} dz_j
  \label{eq:ALE}
\end{equation}
The ALE effects need to be properly centered to serve as a functional decomposition, just as with PDP. ALE has also been implemented in many packages, including \CRANpkg{ALEPlot} \citep{apley2020}.

A method for estimating main effects using SHAP, which will be described later, has also been proposed \citep{iwasawa2023, tan2023}. Let \(\phi_j^\mathrm{shap}(\cdot)\) denote the function that assigns the SHAP value of feature \(j\) to any observation \(\mathbf{x}\). Then, \(f^\mathrm{shap}_j\) is defined as follows:
\begin{equation}
f^{\text{shap}}_j(x_j) = \mathbb{E}{\left[\phi^{\text{shap}}_j(\mathbf{X})\;\middle|\;X_j = x_j\right]}
  \label{eq:SHAPBasedDecomposition}
\end{equation}
However, even when SHAP values are obtained for each observation, estimating this conditional expectation as a smooth function from finite data can be challenging. Therefore, in practical applications, some form of scatterplot smoothing should be applied to the SHAP Dependence Plot \citep{lundberg2018} which is based on
\begin{equation}
\left\{ \left( x_{ij},\ \phi_{ij}^{\mathrm{shap}} \right)\;\middle|\;i = 1, 2, \dots, n \right\}
  \label{eq:SHAPDependencePlot}
\end{equation}

\hypertarget{ceteris-paribus-curve}{%
\subsection*{Ceteris Paribus curve}\label{ceteris-paribus-curve}}
\addcontentsline{toc}{subsection}{Ceteris Paribus curve}

One simple way to analyze the behavior of a prediction function is to vary the values of a single feature while holding other features constant, i.e., ceteris paribus, and observe the changes in predictions. The Ceteris Paribus (CP) curve for an individual prediction \(f(\mathbf{x}_i)\) and feature \(j\) is constructed using a function that shows how the prediction changes when only the value of \(x_j\) is varied, while all other feature values are held constant at their levels in \(\mathbf{x}_i\). This function, denoted as \(\mathrm{CP}_{ij}(x_j)\), is precisely defined as follows:
\begin{equation}
\mathrm{CP}_{ij}(x_j) = f{\left({x}_j, \mathbf{x}_{i,\setminus \{j\}} \right)}
  \label{eq:CP}
\end{equation}
Then, the CP curve is visualized as a line plot representing
\begin{equation}
\left\{ \left( x_j,\ \mathrm{CP}_{ij}(x_j) \right)\;\middle|\;x_j^{\min} \leq x_j \leq x_j^{\max} \right\}
  \label{eq:CPCurve}
\end{equation}
R packages, such as \CRANpkg{ceterisParibus} \citep{ceterisParibus}, provide functionality for generating CP curves.

Individual Conditional Expectation (ICE) plots, introduced in \citet{goldstein2015}, display a collection of CP curves within a single plot. ICE plots are particularly useful for detecting interactions between features, which are indicated by heterogeneity in curve shapes. To better visualize the differences in shape, centered ICE (c-ICE) plots are often used. These plots adjust each CP curve by subtracting its value at a reference point \(x_j^*\), which is typically the minimum observed value, \(x_j^{\min}\):
\begin{equation}
\left\{ \left(x_j,\ \mathrm{CP}_{ij}(x_j) - \mathrm{CP}_{ij}(x_j^*) \right)\;\middle|\;x_j^{\min} \leq x_j \leq x_j^{\max} \right\}
  \label{eq:CenteredICEPlot}
\end{equation}
With all curves aligned to \(0\) at the reference point, differences in their shapes become more apparent. Many packages, like \CRANpkg{ICEbox} \citep{goldstein2015}, offer functionality for visualizing ICE plots.

\hypertarget{additive-feature-attribution-method}{%
\subsection*{Additive feature attribution method}\label{additive-feature-attribution-method}}
\addcontentsline{toc}{subsection}{Additive feature attribution method}

Interpretation methods that break down a predicted value for an individual observation \(f(\mathbf{x}_i)\) into the sum of a baseline \(\phi_{i0}\) and the contributions of \(M\) ``simplified features'' \(\phi_{i1},\dots,\phi_{iM}\) as shown in \eqref{eq:AdditiveFeatureAttribution} are referred to as additive feature attribution methods \citep{lundberg2017}.
\begin{equation}
f(\mathbf{x}_i) \approx g_i{\left(\mathbf{z}^*\right)} = \phi_{i0} + \sum_{j=1}^{M} \phi_{ij}
  \label{eq:AdditiveFeatureAttribution}
\end{equation}
where \(g_i\) is an explanation model of \(f(\mathbf{x}_i)\), defined as follows:
\begin{equation}
g_i{(\mathbf{z})} = \phi_{i0} + \sum_{j=1}^{M} \phi_{ij}\ z_{j}
  \label{eq:ExplanationModel}
\end{equation}
Each simplified feature \(z_j\) in \(\mathbf{z} = (z_1, z_2, \dots, z_M)\) is a binary variable that represents whether the \(j\)-th piece of information among the \(M\) pieces in the observation \(\mathbf{x}_i\) is available. Specifically, \(\mathbf{z}^* = (1, \dots, 1)\) indicates that all information in observation \(\mathbf{x}_i\) is available.

SHAP decomposes a prediction based on the idea of the Shapley value, a game-theoretic approach to contribution decomposition, where the contribution of feature \(j\) in a set of \(d\) features \(D\) is defined as follows:
\begin{equation}
\phi^{\mathrm{shap}}_{ij} = \sum_{S \subseteq D \setminus \{j\}} \frac{|S|! \left(d - |S| - 1 \right)!}{d!}\ \left( v_i(S \cup \{j\}) - v_i(S) \right)
  \label{eq:SHAP}
\end{equation}
where \(v_i(S)\), referred to as the characteristic function, represents the expected model prediction when only the values of features in \(S\) are known. There are multiple ways to define it, each with its advantages and disadvantages \citep{sundararajan2020, aas2021}. The original definition introduced in \citet{lundberg2017} is given as follows:
\begin{equation}
v_i(S) = \mathbb{E} {\left[f{(\mathbf{X})}\;\middle|\;\mathbf{X}_S = \mathbf{x}_{iS} \right]}
  \label{eq:ConditionalExpectationSHAP}
\end{equation}
However, estimating this conditional expectation from empirical data is challenging. Consequently, practical SHAP implementations often approximate or redefine \(v_i(S)\) using the expectation based on the marginal distribution of features not in \(S\) as in \eqref{eq:MarginalExpectationSHAP1}, which is then estimated via \eqref{eq:MarginalExpectationSHAP2}.
\begin{align}
v_i(S) &= \mathbb{E} {\left[ f(\mathbf{x}_{iS}, \mathbf{X}_{\setminus S}) \right]}   \label{eq:MarginalExpectationSHAP1} \\
\hat{v}_i(S) &= \frac{1}{n} \sum_{h=1}^n f(\mathbf{x}_{iS}, \mathbf{x}_{h,{\setminus}S})
  \label{eq:MarginalExpectationSHAP2}
\end{align}
Furthermore, the exact computation of Shapley values is computationally expensive, as it requires \(2^M\) evaluations of \(v_i(S)\) for each observation. To mitigate this challenge, approximation techniques such as Kernel SHAP and model-specific methods like Tree SHAP \citep{lundberg2018} have been developed. Theoretically, additive feature attribution based on Shapley values possesses desirable properties, such as local accuracy and consistency. To visualize SHAP, waterfall plots are commonly used, constructed from the following cumulative sums:
\begin{equation}
\left\{ \left(k,\ \sum_{j \leq k} \phi_{ij}^{\mathrm{shap}} \right)\;\middle|\;k = 0, 1, \dots, M \right\}
  \label{eq:WaterfallPlot}
\end{equation}
Among implementations of SHAP, the Python library \pkg{shap} is the most widely recognized. In R, several implementations are available, including \CRANpkg{shapr} \citep{shapr}, which enables SHAP estimation based on the conditional expectations. The \CRANpkg{shapviz} package \citep{shapviz} is specifically designed for visualizing SHAP values.

LIME \citep{ribeiro2016} is a method for interpreting models by constructing a local surrogate model using predictions in the vicinity of a specific observation, \(\mathbf{x}_i\). In particular, when a linear model is built with simplified features indicating the availability of segmented parts of an image or words in a text, the local model aligns with the definition of an explanation model within the additive feature attribution method. While LIME lacks desirable properties such as local accuracy, it offers advantages in computational efficiency. LIME has several implementations, including the Python library \pkg{lime} \citep{ribeiro2016}.

\hypertarget{feature-importance}{%
\subsection*{Feature importance}\label{feature-importance}}
\addcontentsline{toc}{subsection}{Feature importance}

Methods that summarize the importance of each feature using a real-valued score \(\Phi_j\) are called feature importance methods. Feature importance is often visualized using bar charts based on
\begin{equation}
\left\{ \left(j,\ \Phi_j \right)\;\middle|\;j = 1,2,\dots,d \right\}
  \label{eq:FeatureImportancePlot}
\end{equation}
The \CRANpkg{vip} \citep{vip} package provides a comprehensive set of functions for summarizing and visualizing feature importance.

For ensemble models of classification and regression trees, feature importance \(\Phi_j^\mathrm{gain}\) can be quantified by tracking the total reduction in node impurity (e.g., Gini coefficients) caused by splits for each feature. R packages such as \CRANpkg{randomForest} \citep{randomForest}, \CRANpkg{ranger} \citep{ranger}, and \CRANpkg{xgboost} \citep{chen2016} offer functionality for computing and visualizing this form of feature importance.

Permutation Feature Importance (PFI) is a widely used model-agnostic method for evaluating feature importance. PFI was originally introduced by \citet{breiman2001} for random forest models and later extended into a model-agnostic approach. It quantifies feature importance \(\Phi^\mathrm{pfi}_j\) by measuring the increase in the model's prediction error or deterioration in a performance metric after permuting the values of feature \(x_j\) across the dataset, which breaks the relationships between feature \(j\) and the response variable. The PFI of feature \(j\) is determined by the resulting ratio or difference in losses \citep{fisher2019}.

\citet{friedman2008} proposed the H-statistic as a measure of the strength of interaction effects between features. For a second-order interaction, Friedman's H-statistic can be expressed using the PD-based functional decomposition \eqref{eq:PDBasedDecomposition1} and \eqref{eq:PDBasedDecomposition2} as follows:
\begin{equation}
\hat{\Phi}_{jk}^\mathrm{H^2} = \frac
{\sum_{i = 1}^n{ \hat{f}^\mathrm{pd}_{jk} {\left(x_{ij},x_{ik}\right)}^2}}
{\sum_{i = 1}^n { \left(\hat{f}^\mathrm{pd}_{jk} {\left(x_{ij},x_{ik}\right)} + \hat{f}^\mathrm{pd}_j {\left(x_{ij}\right)} + \hat{f}^\mathrm{pd}_k {\left(x_{ik}\right)} \right) }^2}
  \label{eq:HStatistic}
\end{equation}
The H-statistic does not measure the importance of individual features but instead quantifies the importance of the effects of each term. Several packages, including \CRANpkg{vivid} \citep{vivid}, implement functionality to draw a heatmap based on
\begin{equation}
\left\{ \left(j,\ k,\ \hat{\Phi}_{jk}^\mathrm{H^2} \right)\;\middle|\;(j,\ k) \in {\{1, 2, \dots, d\}}^2,\ j \neq k \right\}
  \label{eq:HStatisticMatrix}
\end{equation}

It is also common to define and estimate feature importance using SHAP values calculated across multiple observations, as described in \eqref{eq:SHAPImportance1} and \eqref{eq:SHAPImportance2}. This approach is known as SHAP Importance.
\begin{align}
\Phi^{\mathrm{shap}}_j &= \mathbb{E}{\left[\left|\phi_j^{\text{shap}}(\mathbf{X})\right|\right]}   \label{eq:SHAPImportance1} \\
\hat{\Phi}^{\mathrm{shap}}_j &= \frac{1}{n} \sum_{i = 1}^n {\left| \phi^\mathrm{shap}_{ij}\right|}
  \label{eq:SHAPImportance2}
\end{align}

\hypertarget{maximum-interpretation-decomposition}{%
\section{Maximum Interpretation Decomposition}\label{maximum-interpretation-decomposition}}

This section provides an overview of Maximum Interpretation Decomposition (MID), a functional decomposition method introduced by \citet{iwasawa2025}. MID serves as a model-agnostic interpretation tool for extracting insights from black-box predictive models.

\hypertarget{definition}{%
\subsection*{Definition}\label{definition}}
\addcontentsline{toc}{subsection}{Definition}

Let \(D = \{1,2,\dots,d\}\) denote the set of \(d\) features, and let \(\mathbf{x}_{i} = (x_{i1},x_{i2},\dots,x_{id})\) represent the feature vector for the \(i\)-th observation, with the target variable denoted as \(y_i\). The prediction function of a predictive model is expressed as \(f\), and its predicted value, \(f(\mathbf{x}_{i})\), is also denoted as \(\hat{y_i}\). The vector obtained by extracting only the features corresponding to a set of features \(J\) from \(\mathbf{x}_i\) is denoted by \(\mathbf{x}_{iJ}\). Where subscripts are not strictly necessary, they may be omitted. When treating these as random variables, they are written as \(\mathbf{X}=(X_1,X_2,\dots,X_d)\), \(\mathbf{X}_J=(X_j)_{j\in J}\) and so on.

A set of functions \(\mathcal{F} = \{f_J\}_{J{\subseteq}D}\), defined for each distinct subsets \(J\), is a \emph{functional decomposition} of \(f\) if it satisfies the following condition:
\begin{equation}
f(\mathbf{x}) = \sum_{J \subseteq D} f_J(\mathbf{x}_J) = f_\emptyset + \sum_j f_{\{j\}}(x_j) + \sum_{j < k} f_{\left\{j, k\right\}}(x_j, x_k) + \dots + f_D(\mathbf{x}_D)
  \label{eq:DefFunctionalDecomposition}
\end{equation}
where each component function \(f_J(\mathbf{x}_J)\) in \eqref{eq:DefFunctionalDecomposition} is referred to as an \emph{effect}, and the \emph{order} of an effect \(f_J\) is defined as the number of features in the feature set \(J\), i.e., \(|J|\). \(f_{\emptyset}\) is referred to as the zero-order effect or intercept, \(f_{\{j\}}\) as the first-order effect or main effect of feature \(X_j\), and \(f_{\{j,k\}}\) as the second-order effect or interaction effect between features \(X_j\) and \(X_k\). Throughout this paper, for simplicity and clarity, \(f_{\{j\}}\) will often be denoted as \(f_j\), \(f_{\{j,k\}}\) as \(f_{jk}\), and so on, when the context makes the feature set clear.

Let \(\mathcal{F}_m\) denote the sum of effects up to order \(m\) included in the functional decomposition \(\mathcal{F}\), expressed as follows:
\begin{equation}
{\mathcal{F}_m}(\mathbf{x})=\sum_{|J|\le m}f_J(\mathbf{x}_J)
  \label{eq:PartialSumOfDecomposition}
\end{equation}
\(\mathcal{F}_m\) can be regarded as a surrogate model that approximately represents the prediction function \(f(\mathbf{x})\). Its representation accuracy is measured using the expected squared error:
\begin{equation}
\mathbb{E} {\left[{(f(\mathbf{X})-{\mathcal{F}_m}(\mathbf{X}))}^2 \right]}
  \label{eq:RepresentationAccuracy}
\end{equation}
In particular, the \emph{uninterpreted variation ratio}, denoted by \(\mathcal{U}\), is defined as follows:
\begin{equation}
\mathcal{U} { \left( {\mathcal{F}_m} \right) } =
{\mathbb{E} { \left[ {(f(\mathbf{X})-{\mathcal{F}_m}(\mathbf{X}))}^2 \right] }}\;\Big{/}\;{\mathbb{E} { \left[ {({f(\mathbf{X})-\mathbb{E} \left[ f(\mathbf{X}) \right] )}^2} \right] }}
  \label{eq:UninterpretedRate}
\end{equation}

For \(m = 1, 2, 3, \dots, d-1\), the \emph{Maximum Interpretation Decomposition of order} \(m\), denoted by \(\mathcal{F}^{\mathrm{mid}(m)}\), is defined as the functional decomposition \(\mathcal{F}\) such that (i) \(\mathcal{F}\) minimizes the uninterpreted variation ratio with respect to the sum of effects up to order \(m\), i.e.,
\begin{equation}
\mathcal{F}^{\mathrm{mid}(m)} = \arg \min_{\mathcal{F}}\ \mathcal{U}{(\mathcal{F}_m)}
  \label{eq:MID}
\end{equation}
(ii) all effects of order \(m+1\) to \(d-1\) are zero, and (iii) every effect of order \(m\) or lower is \emph{strictly centered}. We call the third condition the \emph{strict centering constraint}: an effect \(f_J\) is strictly centered if \(\mathbb{E} {\left[ f_J(\mathbf{X}_J)\;\middle|\;\mathbf{X}_{J'} \right]} = 0\) holds for every proper subset \(J' \subset J\).
The Maximum Interpretation Decomposition of order \(m\) is guaranteed to exist but is not necessarily unique. In cases where multiple candidates for the Maximum Interpretation Decomposition exist, uniqueness with respect to the support of the joint distribution of features can be ensured by introducing the condition that, among the candidates \(\Lambda\), the decomposition that minimizes the expected squared norm is selected, i.e.,
\begin{equation}
\mathcal{F}^{\mathrm{mid}(m)} = \arg \min_{\mathcal{F}\in\Lambda} \sum_{|J|\le m}
\mathbb{E} {\left[ {f_J(\mathbf{X}_J)}^2 \right]}
  \label{eq:MinimumNormCondition}
\end{equation}

Hereafter, the predictive model obtained by summing the effects of order \(m\) or lower in the Maximum Interpretation Decomposition of order \(m\) is referred to as the \emph{MID model of order} \(m\), denoted as \(\mathcal{F}_m^\mathrm{mid}(\mathbf{X})\), and is defined as follows:
\begin{equation}
\mathcal{F}_m^\mathrm{mid}(\mathbf{X}) = \sum_{|J|\leq m} f^{\mathrm{mid}(m)}_J(\mathbf{X}_J), \qquad
f^{\mathrm{mid}(m)}_J \in \mathcal{F}^{\mathrm{mid}(m)}
  \label{eq:MIDModel}
\end{equation}
Where there is no risk of misunderstanding, \(f_J^{\mathrm{mid}(m)}\) is simply written as \(f_J^{\mathrm{mid}}\).

\hypertarget{numerical-estimation}{%
\subsection*{Numerical estimation}\label{numerical-estimation}}
\addcontentsline{toc}{subsection}{Numerical estimation}

For a given observed dataset \((\mathbf{x}_i)_{i=1}^n\) and corresponding model predictions \((\hat{y}_i)_{i=1}^n\), the MID of order \(m\) is estimated by solving the following minimization problem:
\begin{equation}
\hat{\mathcal{F}}^{\mathrm{mid}(m)} = \arg \min_{\mathcal{F}}\sum_{i=1}^n
\left( \hat{y}_i- \sum\nolimits_{|J|\leq m}f_J{\left(\mathbf{x}_{iJ}\right)} \right)^2
  \label{eq:MinimizationProblem}
\end{equation}
subject to the empirical version of the strict centering constraint. That is, for any effect \(f_J\) of order \(1\) to \(m\), any proper subset \(J'\) of \(J\), and any unique observed value \(\mathbf{v}_{J'}\) of \(\mathbf{X}_{J'}\), the following condition holds:
\begin{equation}
\sum_{i\;:\;\mathbf{x}_{iJ'}=\mathbf{v}_{J'}} f_{J}(\mathbf{x}_{iJ}) = 0
  \label{eq:EmpiricalConstraint}
\end{equation}
However, when continuous variables are included among the features, the values of \(\mathbf{x}_{J'}\) typically vary across observations, resulting in the set \(\left\{ i\ :\ \mathbf{x}_{iJ'} = \mathbf{v}_{J'} \right\}\) being a singleton. To ensure the constraint remains meaningful, the equality condition, \(\mathbf{x}_{iJ'} = \mathbf{v}_{J'}\), must be relaxed in some manner. In the implementation of \pkg{midr}, each value of \(\mathbf{x}_{J}\) is represented as a set of weights distributed over a discrete grid, and the constraints are imposed on each grid line. The details are provided in \protect\hyperlink{appendix-numerical-estimation-of-mid}{Appendix}. When the least squares solution to the above problem is not unique, a unique solution is selected by minimizing the empirical sum of squared effects:
\begin{equation}
\hat{\mathcal{F}}^{\mathrm{mid}(m)} = \arg \min_{\mathcal{F}\in\Lambda}
\sum_{|J|\leq m}\sum_{i=1}^n
{f_J{(\mathbf{x}_{iJ})}}^2
  \label{eq:MinimumNormConditionEstim}
\end{equation}
The uninterpreted variation ratio of the MID model \(\hat{\mathcal{F}}^\mathrm{mid}_m\) induced by \(\hat{\mathcal{F}}^{\mathrm{mid}(m)}\) is estimated as follows:
\begin{equation}
\hat{\mathcal{U}} {\left(\hat{\mathcal{F}}^\mathrm{mid}_m \right)} =
{\sum_{i=1}^n {\left(\hat{y}_i-\hat{\mathcal{F}}^\mathrm{mid}_m(\mathbf{x}_i) \right)}^2}\;\Big{/}\;{\sum_{i=1}^n {\left(\hat{y}_i-\bar{\hat{y}} \right)}^2}
  \label{eq:UninterpretedRateEstim}
\end{equation}

\hypertarget{visualization}{%
\subsection*{Visualization}\label{visualization}}
\addcontentsline{toc}{subsection}{Visualization}

For a MID model, the main effect plot for each main effect \(f^\mathrm{mid}_j\) is given by \eqref{eq:MainEffectPlot} and the interaction effect plot for each interaction effect \(f^\mathrm{mid}_{jk}\) is given by \eqref{eq:InteractionPlot}.
In cases where strong correlations between features result in regions without observations (i.e., regions that do not appear in the observed dataset), extrapolation can be avoided by plotting only for points where observations exist:
\begin{equation}
\left\{ \left( x_{ij},\ x_{ik},\ f^\mathrm{mid}_{jk}(x_{ij},x_{ik}) \right)\;\middle|\;
i=1,2,\dots,n\right\}
  \label{eq:EmpiricalInteractionPlot}
\end{equation}

The Ceteris Paribus (CP) curve for feature \(j\) with respect to the MID model \(\mathcal{F}_m^\mathrm{mid}(\mathbf{x})\) is obtained by evaluating \(\mathcal{F}_m^\mathrm{mid}{(x_j,\mathbf{x}_{i,\setminus\{j\}})}\) for a specific instance \(\mathbf{x}_i\) as \(x_j\) varies. This curve can be additively decomposed as shown in \eqref{eq:MIDCP}:
\begin{equation}
\mathrm{CP}^\mathrm{mid}_{ij}(x_j) =
\mathcal{F}^\mathrm{mid}_m {\left(x_j,\mathbf{x}_{i,{\setminus}\{j\}} \right)} =
\sum_{J:j\in J} f^\mathrm{mid}_J {\left(x_j,\mathbf{x}_{i,J{\setminus}\{j\}} \right)}\ +
\sum_{J:j\notin J} f^\mathrm{mid}_J {(\mathbf{x}_{iJ})}
  \label{eq:MIDCP}
\end{equation}
and it can be visualized based on \eqref{eq:CPCurve}. Furthermore, it is possible to decompose the contribution of each second-order interaction effect to the CP curves by creating line plots representing
\begin{equation}
\left\{ \left(x_j,\ f^\mathrm{mid}_{jk}{(x_j,x_{ik})} \right)\;\middle|\;x_j^{\min}\ {\le}\ x_j\ {\le}\ x_j^{\max} \right\}
  \label{eq:MIDEffectCP}
\end{equation}

The prediction of a MID model for an observation \(\mathbf{x}_i\) can be expressed as the sum of the intercept and the effects of order \(1\) to \(m\) as follows:
\begin{equation}
\mathcal{F}^\mathrm{mid}_m(\mathbf{x}_i) = f^\mathrm{mid}_\emptyset + \sum_{|J|=1,\dots,m}f^\mathrm{mid}_J(\mathbf{x}_{iJ})
  \label{eq:MIDPrediction}
\end{equation}
Due to the strict centering constraint of MID, the conditional expectation of the effect \(f^\mathrm{mid}_J(\mathbf{X}_J)\), given \(\mathbf{X}_{J'} = \mathbf{x}_{i{J'}}\) for any \(J' \subseteq J\), is expressed as follows:
\begin{equation}
\mathbb{E}\left[f^\mathrm{mid}_J(\mathbf{X}_J)\;\middle|\;\mathbf{X}_{J'} = \mathbf{x}_{i{J'}} \right] = \begin{cases} f^\mathrm{mid}_J(\mathbf{x}_{iJ}) & \text{if}\quad {J'} = J \\
0 & \text{if}\quad {J'} \subset J \end{cases}
  \label{eq:MIDEffect}
\end{equation}
Thus, \(f^\mathrm{mid}_J(\mathbf{x}_{iJ})\) can be viewed as the contribution of the feature set \(J\) to the prediction. This perspective allows us to regard MID as an additive attribution method, analogous to \eqref{eq:AdditiveFeatureAttribution}. To formalize this, let \(z_J\) in \(\mathbf{z} = (z_J)_{|J| = 1,\dots,m}\) be a binary indicator for the full availability of features in \(J\). Then, \(\mathcal{F}^\mathrm{mid}_m(\mathbf{x}_i)\) can be expressed using an explanation model \(g_i(\mathbf{z})=\phi_{\emptyset}^\mathrm{mid}+\sum_{|J|=1,\dots,m}\phi_{iJ}^\mathrm{mid}z_J\) where \(\phi^\mathrm{mid}_\emptyset = f^\mathrm{mid}_\emptyset\) and \(\phi^\mathrm{mid}_{iJ} = f^\mathrm{mid}_J(\mathbf{x}_{iJ})\). When all effects up to order \(m\) are available (i.e., when \(z_J = 1\) for all relevant \(J\), denoted as \(\mathbf{z}^*\)), the explanation model \(g_i\) recovers the MID prediction:
\begin{equation}
f(\mathbf{x}_i) \approx \mathcal{F}^\mathrm{mid}_m(\mathbf{x}_i) =
g_i(\mathbf{z}^*) =
\phi^\mathrm{mid}_\emptyset + \sum_{|J|=1,\dots,m} \phi^\mathrm{mid}_{iJ}
  \label{eq:MIDBreakDown}
\end{equation}
By appropriately sorting \(\phi^\mathrm{mid}_{iJ}\), a waterfall plot based on cumulative sums \eqref{eq:WaterfallPlot} can be generated. Furthermore, feature-level attributions for the MID model predictions can be calculated exactly and efficiently using the SHAP framework. This is achieved by first defining the characteristic function \(v_i(S)\) for the MID model as follows:
\begin{equation}
v_i(S) = \sum_{|{J}| \leq m} \mathbb{E} {\left[ f^\mathrm{mid}_{J} {(\mathbf{X}_{J})}\;\middle|\;\mathbf{X}_{S\cap {J}} = \mathbf{x}_{i,S\cap {J}} \right]} = \sum_{{J}\subseteq S,\ |J| \leq m} f^\mathrm{mid}_{J} {(\mathbf{x}_{i{J}})}
  \label{eq:MIDSHAPv}
\end{equation}
Applying \eqref{eq:SHAP} to this specific \(v_i(S)\) yields a closed-form solution for the MID-derived Shapley values, denoted as \(\phi_{ij}^\mathrm{ms}\), which is given as follows:
\begin{equation}
\phi^\mathrm{ms}_{ij} = \sum_{J:j\in J,\ |J|\leq m} \frac{1}{|J|} {f^\mathrm{mid}_J {(\mathbf{x}_{iJ})}}
  \label{eq:MIDSHAP}
\end{equation}

As shown in \eqref{eq:MIDEffect}, \(f^\mathrm{mid}_J (\mathbf{X}_J)\) can be regarded as a random variable that represents the contribution of the full availability of all features in \(J\) to prediction. In light of this, the importance of each effect \(f^\mathrm{mid}_J\) can be defined and estimated as in \eqref{eq:MIDEffectImportance1} and \eqref{eq:MIDEffectImportance2}, and can be visualized based on \eqref{eq:FeatureImportancePlot}.
\begin{align}
{\Phi}^\mathrm{mid}_J &= \mathbb{E}
{\left[ \left| f^\mathrm{mid}_J(\mathbf{X}_{J}) \right| \right]}
  \label{eq:MIDEffectImportance1} \\
\hat{\Phi}^\mathrm{mid}_J &= \frac{1}{n}{{\sum}_{i=1}^n{\left|{f}^\mathrm{mid}_J{(\mathbf{x}_{iJ})}\right|}}
  \label{eq:MIDEffectImportance2}
\end{align}
Alternatively, the importance of feature \(j\) can be calculated using \(\phi^\mathrm{ms}_{ij}\) in \eqref{eq:SHAPImportance2}.

\hypertarget{constructing-mid-in-r}{%
\section{Constructing MID in R}\label{constructing-mid-in-r}}

\begin{figure}
\centering
\includegraphics{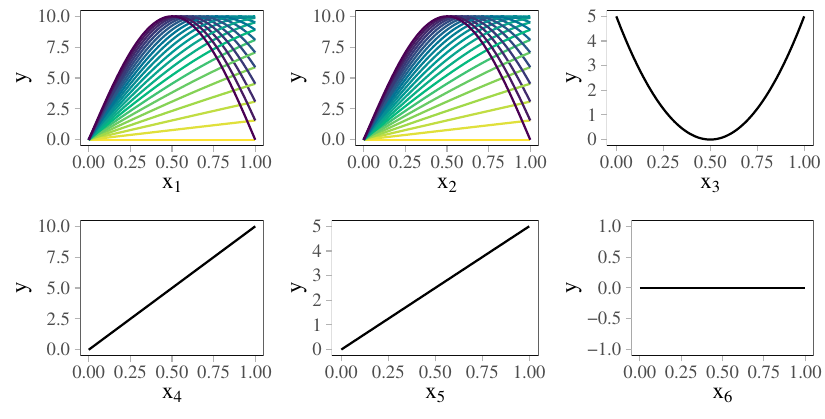}
\caption{\label{fig:friedman}True effects of features from \(x_1\) (\emph{top-left}) to \(x_6\) (\emph{bottom-right}). The interaction effect between \(x_1\) and \(x_2\) is represented by colored lines.}
\end{figure}

In this section, we demonstrate how to interpret a regression model using the \pkg{midr} package in R. For this purpose, we perform a numerical simulation based on a regression problem described in \citet{friedman1991}. The training dataset contains 10 independent features \(x_1,x_2,\dots,x_{10}\), each uniformly distributed over the interval \(\left[0, 1 \right]\), and a target variable \(y\) generated according to the following equation with the disturbance term \(\epsilon\ {\sim}\ \mathcal{N}{(0,1)}\).
\begin{equation}
y=10\sin{(\pi{x_1}{x_2})+20{(x_3-0.5)}^2+10x_4+5x_5+\epsilon}
  \label{eq:FriedmanBenchmark}
\end{equation}
Figure \ref{fig:friedman} illustrates the effects of each feature on the response variable. The interaction effect between \(x_1\) and \(x_2\) is represented by colored lines: the effect of \(x_1\) depends on the value of \(x_2\) (yellow for \(x_2 = 0\) and purple for \(x_2 = 1\)), and vice versa.

In the simulation, both the training and testing datasets consist of 2,000 samples.

\begin{verbatim}
library(mlbench)
set.seed(42)
train <- as.data.frame(mlbench.friedman1(n = 2000L))
test <- as.data.frame(mlbench.friedman1(n = 2000L))
\end{verbatim}

As the target model for interpretation, a neural network model is built with the \CRANpkg{nnet} \citep{nnet} package.

\begin{verbatim}
library(nnet)
model <- nnet(y ~ ., train, size = 10, linout = TRUE, maxit = 1e3, decay = 0.1)
preds <- predict(model, test)[, 1]
sqrt(mean((preds - test$y) ^ 2)) # RMSE
\end{verbatim}

\begin{verbatim}
#> [1] 1.07446
\end{verbatim}

To construct a MID model of the prediction function, the \texttt{interpret()} function is used. The \texttt{formula} argument to \texttt{interpret()} specifies the terms to be included in the decomposition. In the following code, the formula is set to include all first-order main effects and second-order interactions. The return value is an object with the class attribute ``mid''.

\begin{verbatim}
library(midr)
mid <- interpret(y ~ .^2, data = train, model = model)
mid
\end{verbatim}

\begin{verbatim}
#>
#> Call:
#> interpret(formula = yhat ~ .^2, data = train, model = model)
#>
#> Model Class: nnet.formula, nnet
#>
#> Intercept: 14.234
#>
#> Main Effects:
#> 10 main effect terms
#>
#> Interactions:
#> 45 interaction terms
#>
#> Uninterpreted Variation Ratio: 0.0011647
\end{verbatim}

\noindent The uninterpreted variation ratio is notably low, indicating that the MID model's prediction function accurately approximates that of the neural network model on the training dataset.

\begin{figure}
\centering
\includegraphics{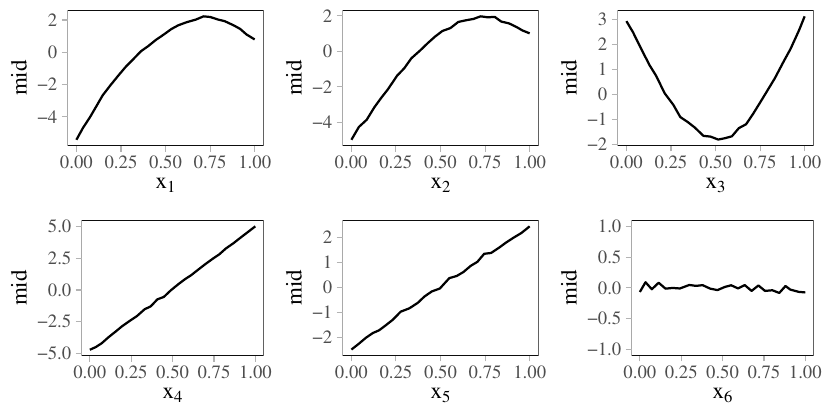}
\caption{\label{fig:maineffects}Main effect plots displaying the feature effects from \(x_1\) (\emph{top-left}) to \(x_6\) (\emph{bottom-right}).}
\end{figure}

For a ``mid'' object, two types of graphing functions, \texttt{plot()} and \texttt{ggmid()}, can be used to generate main effect plots \eqref{eq:MainEffectPlot} and interaction effect plots \eqref{eq:InteractionPlot}. These plots are created by specifying the \texttt{term} to be visualized, such as \texttt{plot(mid,\ term\ =\ "x.1")} or \texttt{ggmid(mid,\ term\ =\ "x.1")}. The former utilizes standard packages such as \CRANpkg{graphics}, while the latter relies on \CRANpkg{ggplot2} \citep{ggplot2}. Figure \ref{fig:maineffects} presents main effect plots for six features from \(x_1\) to \(x_6\).

\begin{figure}
\centering
\includegraphics{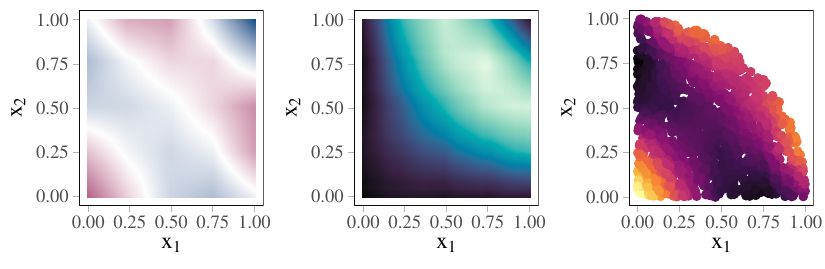}
\caption{\label{fig:interactions}Interaction plots. \emph{Left}: Simple interaction effect plot. \emph{Middle}: Sum of main and interaction effects. \emph{Right}: Interaction effect plot based on limited data points.}
\end{figure}

Figure \ref{fig:interactions} presents examples of interaction effect plots. The inclusion of main effects can be controlled using the \texttt{main.effects} argument of the graphing functions: the left plot shows \(f^\mathrm{mid}_{1, 2}(x_1, x_2)\) while the middle plot displays \(f^\mathrm{mid}_{1, 2}(x_1, x_2) + f^\mathrm{mid}_{1}(x_1) + f^\mathrm{mid}_{2}(x_2)\). As shown in the right plot, specifying \texttt{type\ =\ "data"} allows plotting only the values at points where observations exist as described in \eqref{eq:EmpiricalInteractionPlot}. The appearance of the plot can be easily customized by specifying a color theme name in the \texttt{theme} argument.

\begin{verbatim}
ggmid(mid, term = "x.1:x.2") # Figure 3, Left
ggmid(mid, term = "x.1:x.2", theme = "Mako", main.effects = TRUE) # Middle
ggmid(mid, term = "x.1:x.2", type = "data", theme = "Inferno",
      data = train[train$x.1^2 + train$x.2^2 < 1,]) # Right
\end{verbatim}

\begin{figure}
\centering
\includegraphics{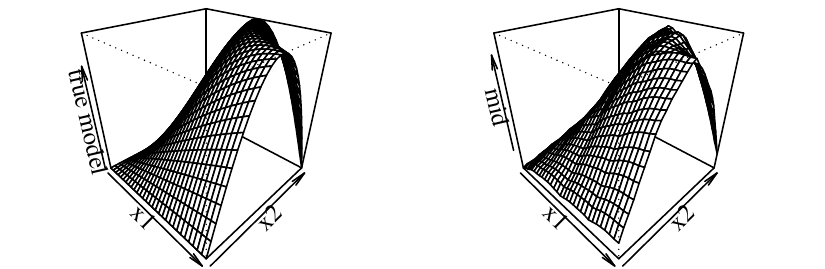}
\caption{\label{fig:interaction3d}Comparison of 3D interaction plots. \emph{Left}: True function. \emph{Right}: Sum of the main and interaction effects of the fitted MID model.}
\end{figure}

Figure \ref{fig:interaction3d} compares the true function with the sum of corresponding effects of the MID model, \(f^\mathrm{mid}_{1,2}(x_1,x_2) + f^\mathrm{mid}_1(x_1) + f^\mathrm{mid}_2(x_2)\), derived by interpreting the \pkg{nnet} model. As can be observed from the figure, the MID model effectively approximates the shape of the true function.

\begin{figure}
\centering
\includegraphics{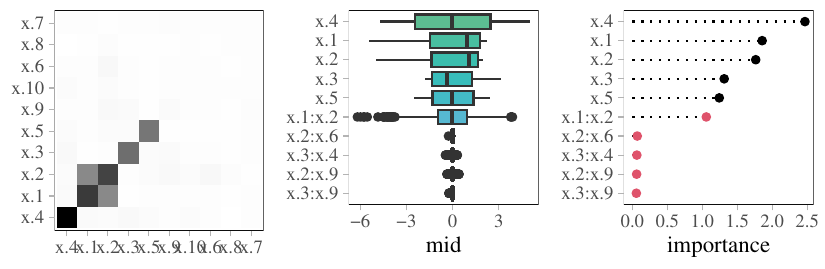}
\caption{\label{fig:importance}Feature importance plots. \emph{Left}: Heatmap. \emph{Middle}: Boxplot. \emph{Right}: Cleveland dot plot.}
\end{figure}

By passing a ``mid'' object to the \texttt{mid.importance()} function, a ``mid.importance'' object is created that summarizes the effect importance of the fitted MID model, which can be visualized using the graphing functions. As shown in Figure \ref{fig:importance}, specifying the \texttt{type} argument allows for generating various types of visualizations, including bar plots or dot charts based on \eqref{eq:FeatureImportancePlot}, heatmaps like \eqref{eq:HStatisticMatrix}, and box plots representing the distribution of effect sizes.

\begin{verbatim}
imp <- mid.importance(mid)
ggmid(imp, type = "heatmap") # Figure 5, Left
ggmid(imp, type = "boxplot", theme = "Cold", max.bars = 10) # Middle
ggmid(imp, type = "dotchart", theme = "R4", max.bars = 10) # Right
\end{verbatim}

\begin{figure}
\centering
\includegraphics{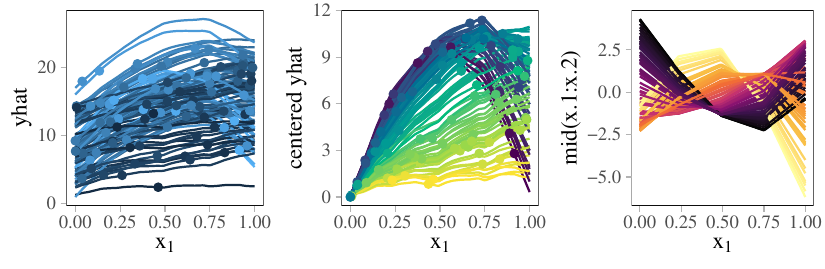}
\caption{\label{fig:iceplot}ICE plots for \(x_1\). \emph{Left}: ICE plot. \emph{Middle}: Centered ICE (c-ICE) plot. \emph{Right}: ICE for the interaction effect between \(x_1\) and \(x_2\).}
\end{figure}

By passing a ``mid'' object and the reference dataset to the \texttt{mid.conditional()} function, a ``mid.conditional'' object for generating ICE plots \eqref{eq:CPCurve} is created. As shown in Figure \ref{fig:iceplot}, setting the \texttt{type} argument in the graphing functions to ``centered'' enables the creation of c-ICE plots \eqref{eq:CenteredICEPlot}. It is also possible to generate an ICE plot showing how the specified effect term varies across observations based on the idea in \eqref{eq:MIDEffectCP} by specifying the \texttt{term} argument.

\begin{verbatim}
ice <- mid.conditional(mid, variable = "x.1", data = train[1:100, ])
ggmid(ice, var.color = "x.2") # Figure 6, Left
ggmid(ice, type = "centered", theme = "Viridis_r", var.color = "x.2") # Middle
ggmid(ice, term = "x.1:x.2", theme = "Inferno", var.color = "x.2", dots = FALSE) # Right
\end{verbatim}

\begin{figure}
\centering
\includegraphics{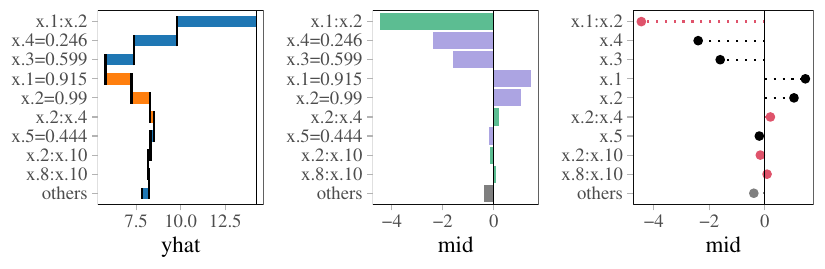}
\caption{\label{fig:breakdown}Prediction breakdown plots. \emph{Left}: Waterfall plot. \emph{Middle} Barplot. \emph{Right}: Cleveland dot plot with simple axis labels.}
\end{figure}

By passing a ``mid'' object and a single observation to the \texttt{mid.breakdown()} function, a ``mid.breakdown'' object is created. With the graphing functions, it enables one to generate plots that decompose the MID model's prediction into the contributions of each effect term. As shown in Figure \ref{fig:breakdown}, waterfall plots like \eqref{eq:WaterfallPlot} can be generated, as well as bar plots and dot charts.

\begin{verbatim}
pbd <- mid.breakdown(mid, data = train[1, ], digits = 3)
ggmid(pbd, theme = "Classic Tableau", max.bars = 10) # Figure 7, Left
ggmid(pbd, type = "barplot", theme = "Cold", max.bars = 10) # Middle
ggmid(pbd, type = "dotchart", theme = "R4", max.bars = 10, format = "%t") # Right
\end{verbatim}

With the \pkg{shapviz} package, the plots based on the MID-derived Shapley values \(\phi_{ij}^{ms}\), defined by \eqref{eq:MIDSHAP}, can be generated. By passing a ``mid'' object along with the reference dataset to the \texttt{shapviz()} function, a ``shapviz'' object based on \(\phi_{ij}^{ms}\) is quickly constructed. This object is compatible with visualization functions in the \pkg{shapviz} package, including \texttt{sv\_dependence()} for SHAP dependence plots \eqref{eq:SHAPDependencePlot} and \texttt{sv\_importance()} for SHAP feature importance plots based on \eqref{eq:SHAPImportance2}.

\begin{verbatim}
library(shapviz)
spv <- shapviz(mid, data = train)
sv_dependence(spv, "x.1") # Figure 8, Left
sv_dependence2D(spv, "x.1", "x.2", viridis_args = list(option = "mako")) # Middle
sv_importance(spv, fill = "gray25") # Right
\end{verbatim}

\begin{figure}
\centering
\includegraphics{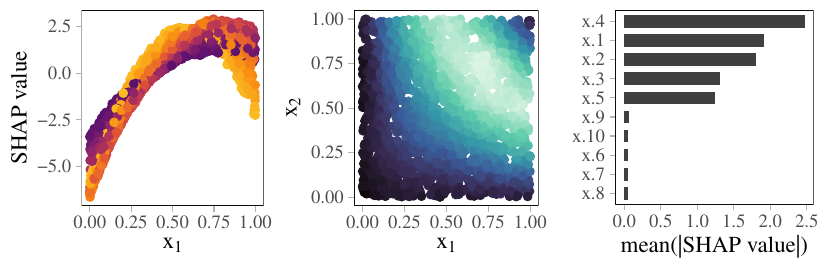}
\caption{\label{fig:shap}MID-derived SHAP plots. \emph{Left}: SHAP dependence plot. \emph{Middle}: SHAP 2D dependence plot. \emph{Right}: SHAP importance plot.}
\end{figure}

It should be noted that, as an interpretation method, \pkg{midr} summarizes and visualizes the interpretative surrogate model \(\mathcal{F}^\mathrm{mid}_m(\mathbf{X})\), which represents the target model \(f(\mathbf{X})\), in various ways. For \(\mathcal{F}_m^\mathrm{mid}\) to be considered a good representation of the target model, it is essential to verify that it accurately represents the target model. While the uninterpreted variation ratio can be one important criterion, a low uninterpreted variation ratio on the training dataset alone does not guarantee that the target model and the interpretative MID model will produce similarly close predictions for new observations. In other words, the interpretative MID model might be overfitted to the target model's predictions on the training dataset. As an example of verification, we can calculate the uninterpreted variation ratio based on the testing dataset.

\begin{verbatim}
sum((preds - predict(mid, test))^2) / sum((preds - mean(preds))^2)
\end{verbatim}

\begin{verbatim}
#> [1] 0.005303483
\end{verbatim}

For more detailed examples of interpreting regression and classification models using the \pkg{midr} package, please explore the vignettes provided on the website \url{https://ryo-asashi.github.io/midr/}.

\hypertarget{discussion}{%
\section{Discussion}\label{discussion}}

This section discusses two key characteristics of MID as an interpretation method, pragmatic stability and computational efficiency, based on numerical simulations.

\hypertarget{pragmatic-stability}{%
\subsection*{Pragmatic stability}\label{pragmatic-stability}}
\addcontentsline{toc}{subsection}{Pragmatic stability}

Consider two prediction functions, \(f_A\) and \(f_B\), satisfying \(\mathrm{P} {\left( f_A(\mathbf{X}) = f_B(\mathbf{X}) \right)} = 1\). These two functions are empirically equivalent, consistently producing identical predictions for the combinations of features observable in the real world. A functional decomposition method that assigns the same decomposition to empirically equivalent prediction functions can be called \emph{pragmatically stable}.

\begin{figure}
\centering
\includegraphics{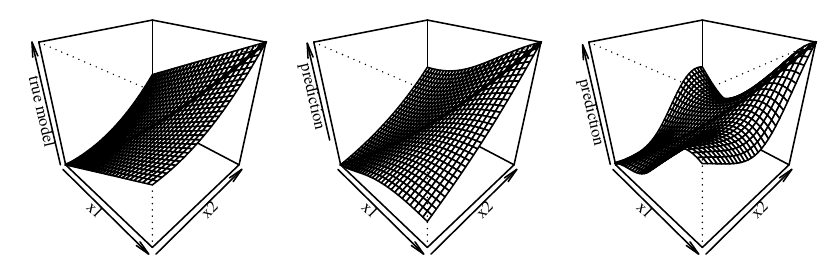}
\caption{\label{fig:comparison}Comparison of interaction effects of the true model and newral network models. \emph{Left}: True function. \emph{Middle} and \emph{Right}: Fitted prediction functions of neural network models.}
\end{figure}

To examine the pragmatic stability of MID, we performed a numerical simulation described in \citet{apley2020}. Let \((X_1, X_2)\) be a feature vector defined as \((X_1, X_2) = (U + Z_1, U + Z_2)\), where \(U\) is a uniform random variable on \(\left[ 0, 1 \right]\), and \(Z_1\) and \(Z_2\) are independent normal random variables following \(\mathcal{N}{(0, {0.05}^2)}\), and the target variable \(Y\) is computed by \(Y = X_1 + X_2^2 + \epsilon\), where \(\epsilon\) is a disturbance term following \(\mathcal{N}{(0, {0.1}^2)}\). A dataset, consisting of \(200\) observations generated from this distribution, is used to train a single hidden layer neural network model using the \pkg{nnet} package. Since the shape of the prediction function is influenced by the observations and the initial values of the model parameters, a Monte Carlo simulation with 50 trials was conducted. Figure \ref{fig:comparison} compares the true function \(f(x_1,x_2) = x_1 + {x_2}^2\) with two prediction functions of the fitted neural network models. Within the region of observations, particularly where \(x_1 \approx x_2\), the prediction functions closely align with the true function. However, in areas where no observations exist, the \pkg{nnet} models do not necessarily approximate the true function.

\begin{figure}
\centering
\includegraphics{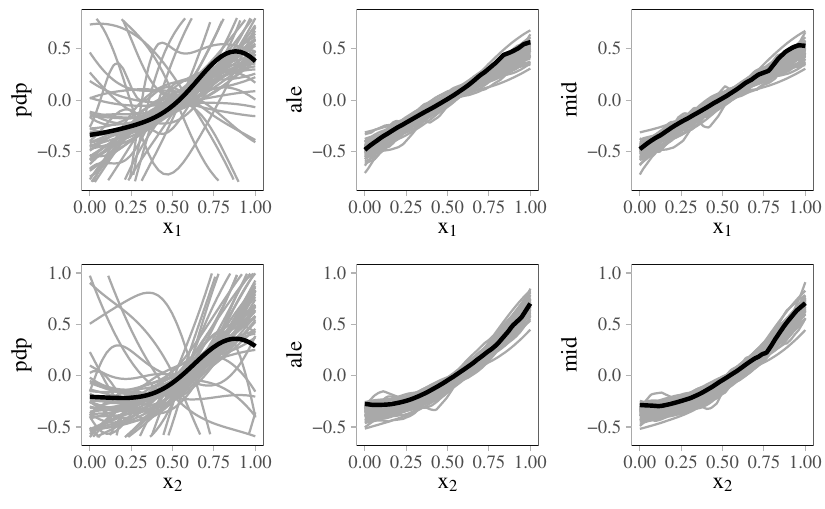}
\caption{\label{fig:montecarlo}Comparison of main effect plots. \emph{Left}: PDP. \emph{Middle}: ALE. \emph{Right}: MID.}
\end{figure}

Figure \ref{fig:montecarlo} illustrates the main effects of the fitted neural networks, estimated using PDP, ALE, and MID. The estimated effects across 49 trials are displayed in gray, while the result of the 50th trial is shown in black. As shown in Figure \ref{fig:montecarlo}, the PD-based main effect plots are influenced by extrapolations beyond the envelope of the observed dataset, resulting in a divergence from the true functions. This indicates that in the presence of feature correlations, PDP lacks pragmatic stability and may yield invalid interpretations. In contrast, across all trials of this simulation, MID closely aligns with ALE, consistently producing main effect plots that accurately reflect the true function.

\hypertarget{computational-efficiency}{%
\subsection*{Computational efficiency}\label{computational-efficiency}}
\addcontentsline{toc}{subsection}{Computational efficiency}

In the implementation of the \texttt{interpret()} function, the \texttt{fastLmPure()} function from the \CRANpkg{RcppEigen} \citep{RcppEigen} package is employed to solve least squares problems. The \texttt{method} argument of \texttt{interpret()} is directly passed to the corresponding argument in \texttt{fastLmPure()}. The default value of \texttt{method} is determined by \texttt{singular.ok}, which specifies whether a non-unique least squares solution arising from rank deficiency is permitted. If rank deficiency is permitted (\texttt{singular.ok = TRUE}), a least squares solution is obtained using eigendecomposition (\texttt{method = 5}), and if it is not allowed, the column-pivoted QR decomposition (\texttt{method = 0}) is used. Both methods support rank determination. Notably, the eigendecomposition approach outputs a minimum-norm least squares solution when the design matrix is rank-deficient, which ensures the uniqueness of MID. Additionally, alternative numerical solution methods can be utilized by specifying different values for the \texttt{method} argument in the \texttt{interpret()} function. For instance, adopting the \(\mathrm{LL^T}\) Cholesky decomposition method (\texttt{method = 2}) can lead to performance improvements; however, this comes at the cost of losing the ability to determine rank deficiency.

In the numerical estimation of MID described in \protect\hyperlink{appendix-numerical-estimation-of-mid}{Appendix}, the computational cost of solving the least squares problem becomes dominant as the problem size increases. Consequently, the choice of the \texttt{method} can have a significant impact on overall computation time. Table \ref{tab:evaltime} presents the results of a simulation performed to compare the computation times across various numerical least squares algorithms. The simulation employed the \CRANpkg{mlbench} \citep{mlbench} package along with the \CRANpkg{microbenchmark} \citep{microbenchmark} package, and the essence of the simulation method is outlined in the following code:

\begin{verbatim}
circle <- as.data.frame(mlbench.circle(n = n, d = d))
microbenchmark(interpret(classes ~ .^2, data = circle, k = c(25, 5), method = method))
\end{verbatim}

\noindent where the \texttt{k} argument in \texttt{interpret()} specifies the number of encoding functions used for each feature in modeling main effects (\(k_{(1)} = 25\)) and for each feature comprising second-order interactions (\(k_{(2)} = 5\)), respectively, and the \texttt{method} argument specifies the least squares algorithm to be used.

In Table \ref{tab:evaltime}, \(n\) and \(d\) denote the number of rows and columns of the dataset, respectively, while \(m\) represents the number of parameters estimated in the least squares method. Since our model in this simulation includes all possible second-order effects, the value of \(m\) is given by \(m = k_{(1)}d + k_{(2)}^2d(d-1)/2\) where \(k_{(1)} = 25\) and \(k_{(2)} = 5\), which therefore grows quadratically with \(d\). The subsequent columns in the table, each corresponding to a numerical least squares algorithm, report computation times in milliseconds. For small \(m\), the overhead from preprocessing the design matrix and postprocessing the results, which have a time complexity of \(O(nm)\), cannot be ignored. Consequently, computation time generally increases proportionally with the number of observations \(n\). However, as \(m\) becomes larger, the differences in computation times across least square methods become increasingly significant. For the case with \(n = 10,000\) observations and \(m = 3,400\) parameters derived from \(d = 16\) features, the method utilizing Jacobi SVD (\texttt{method = 4}) requires significantly more computation time than other methods. Meanwhile, the method using \(\mathrm{LL^T}\) Cholesky decomposition achieves computation times that are approximately \(35\)\% to \(40\)\% of those for the column-pivoted QR decomposition.

\begin{longtable}[]{@{}
  >{\raggedleft\arraybackslash}p{(\columnwidth - 16\tabcolsep) * \real{0.0976}}
  >{\raggedleft\arraybackslash}p{(\columnwidth - 16\tabcolsep) * \real{0.0488}}
  >{\raggedleft\arraybackslash}p{(\columnwidth - 16\tabcolsep) * \real{0.0732}}
  >{\raggedleft\arraybackslash}p{(\columnwidth - 16\tabcolsep) * \real{0.1220}}
  >{\raggedleft\arraybackslash}p{(\columnwidth - 16\tabcolsep) * \real{0.1220}}
  >{\raggedleft\arraybackslash}p{(\columnwidth - 16\tabcolsep) * \real{0.1220}}
  >{\raggedleft\arraybackslash}p{(\columnwidth - 16\tabcolsep) * \real{0.1220}}
  >{\raggedleft\arraybackslash}p{(\columnwidth - 16\tabcolsep) * \real{0.1463}}
  >{\raggedleft\arraybackslash}p{(\columnwidth - 16\tabcolsep) * \real{0.1463}}@{}}
\caption{\label{tab:evaltime} Average evaluation time in milliseconds of \texttt{interpret()} for various task sizes and least squares algorithms from \texttt{fastLmPure()}.}\tabularnewline
\toprule\noalign{}
\begin{minipage}[b]{\linewidth}\raggedleft
n
\end{minipage} & \begin{minipage}[b]{\linewidth}\raggedleft
d
\end{minipage} & \begin{minipage}[b]{\linewidth}\raggedleft
m
\end{minipage} & \begin{minipage}[b]{\linewidth}\raggedleft
pivoted QR
\end{minipage} & \begin{minipage}[b]{\linewidth}\raggedleft
unpivoted QR
\end{minipage} & \begin{minipage}[b]{\linewidth}\raggedleft
\(\mathrm{LL^T}\) Cholesky
\end{minipage} & \begin{minipage}[b]{\linewidth}\raggedleft
\(\mathrm{LDL^T}\) Cholesky
\end{minipage} & \begin{minipage}[b]{\linewidth}\raggedleft
Jacobi SVD
\end{minipage} & \begin{minipage}[b]{\linewidth}\raggedleft
eigenvalue
\end{minipage} \\
\midrule\noalign{}
\endfirsthead
\toprule\noalign{}
\begin{minipage}[b]{\linewidth}\raggedleft
n
\end{minipage} & \begin{minipage}[b]{\linewidth}\raggedleft
d
\end{minipage} & \begin{minipage}[b]{\linewidth}\raggedleft
m
\end{minipage} & \begin{minipage}[b]{\linewidth}\raggedleft
pivoted QR
\end{minipage} & \begin{minipage}[b]{\linewidth}\raggedleft
unpivoted QR
\end{minipage} & \begin{minipage}[b]{\linewidth}\raggedleft
\(\mathrm{LL^T}\) Cholesky
\end{minipage} & \begin{minipage}[b]{\linewidth}\raggedleft
\(\mathrm{LDL^T}\) Cholesky
\end{minipage} & \begin{minipage}[b]{\linewidth}\raggedleft
Jacobi SVD
\end{minipage} & \begin{minipage}[b]{\linewidth}\raggedleft
eigenvalue
\end{minipage} \\
\midrule\noalign{}
\endhead
\bottomrule\noalign{}
\endlastfoot
1,000 & 2 & 75 & 54.3 & 50.1 & 48.2 & 43.2 & 100.6 & 45.8 \\
10,000 & 2 & 75 & 289.8 & 275.9 & 243.3 & 222.0 & 540.3 & 244.8 \\
100,000 & 2 & 75 & 2,468.0 & 2,424.9 & 2,065.7 & 2,011.8 & 5,259.3 & 2,159.4 \\
10,000 & 4 & 250 & 680.2 & 656.8 & 560.6 & 561.7 & 3,533.5 & 669.2 \\
10,000 & 8 & 900 & 4,981.0 & 3,029.2 & 2,451.5 & 2,369.4 & 52,360.1 & 3,640.3 \\
10,000 & 16 & 3,400 & 50,178.0 & 23,204.4 & 17,582.8 & 19,667.1 & 1,829,342.0 & 54,829.4 \\
\end{longtable}

\hypertarget{summary}{%
\section{Summary}\label{summary}}

We have introduced the R package \pkg{midr}, a tool for interpreting black-box models by constructing a global surrogate model based on the functional decomposition of the target model's prediction function.

Functional decomposition is considered the ``key concept of machine learning interpretability'' \citep{molnar2025}. By explicitly considering interaction effects, it enables the unification of local and global explanations \citep{hiabu2022}. As a tool for global interpretation, additive representations of predictive models obtained by functional decomposition can outperform other types of representations across various interpretation tasks \citep{tan2023}. We believe further research on functional decomposition is necessary to advance our understanding of model interpretability.

By definition, the MID model of order \(m\) is the most accurate global additive representation among additive representations consisting of terms of at most order \(m\). Furthermore, due to the constraints imposed on it, it possesses a wide range of analytical capabilities. As an implementation of MID, \pkg{midr} offers a comprehensive set of functions for learning from black-box models.

\hypertarget{acknowledgements}{%
\section{Acknowledgements}\label{acknowledgements}}

The development of the \pkg{midr} package was conducted as part of the activities of the Data Science Related Basic Research Subcommittee of the Institute of Actuaries of Japan (IAJ).

\hypertarget{appendix-numerical-estimation-of-mid}{%
\section*{\texorpdfstring{Appendix \quad Numerical estimation of MID}{Appendix Numerical estimation of MID}}\label{appendix-numerical-estimation-of-mid}}
\addcontentsline{toc}{section}{Appendix \quad Numerical estimation of MID}

This appendix outlines the numerical estimation method for MID, as implemented in the \pkg{midr} package. The method assumes that the observed dataset is structured as a matrix with \(n\) rows and \(d\) columns, accompanied by an \(n\)-dimensional vector of model predictions, \({\hat{\mathbf{y}}}={(\hat{y}_1,\dots,\hat{y}_n)}^\mathrm{T}\), corresponding to these observations.

The intercept of MID is the expected value of the model prediction and can be estimated as follows:
\begin{equation}
\hat{f}^\mathrm{mid}_{\emptyset} = \frac{1}{n} \sum_{i=1}^n \hat{y_i}
  \label{eq:Appendix1}
\end{equation}
Subsequently, the \(n\)-dimensional vector obtained by subtracting the intercept from the predicted values is expressed as \(\tilde{\mathbf{y}} = \hat{\mathbf{y}} - \hat{f}^\mathrm{mid}_{\emptyset}\mathbf{1}\), where \(\mathbf{1}\) is an \(n\)-dimensional vector of ones.

\hypertarget{first-order-mid}{%
\subsection*{First-order MID}\label{first-order-mid}}
\addcontentsline{toc}{subsection}{First-order MID}

The main effect \(f_j(x_j)\) of feature \(j\) is modeled using \(k_j\) parameters \(\{\beta_j^s\}_{s=1}^{k_j}\) and \(k_j\) encoding functions \(\{\chi_j^s\}_{s=1}^{k_j}\), which satisfy \(\sum_{s=1}^{k_j}\chi_j^s(x_j)=1\) for any \(x_j\), as follows:
\begin{equation}
f_{j}(x_j) = \sum_{s=1}^{k_j} \beta_j^s \chi_j^s(x_j)
  \label{eq:Appendix2}
\end{equation}
Let \(k'_j\) be the number of unique values of feature \(x_j\). If \(k'_j\) is finite and sufficiently small, we set \(k_j = k'_j\). Let the unique values be \(x_j^{(1)}, x_j^{(2)}, \dots, x_j^{(k_j)}\). The encoding functions \(\chi_{j}^{s}(x_j)\) are then defined as indicator functions:
\begin{equation}
\chi_{j}^{s}(x_j) = \begin{cases}
1 \quad & \mathrm{if}\quad x_j = x_j^{(s)} \\
0 & \mathrm{if}\quad x_j \neq x_j^{(s)}
\end{cases},\quad s = 1,\dots,{k_j}
  \label{eq:Appendix3}
\end{equation}
If \(k'_j\) is infinite, or if finite but too large, \(k_j\) is selected as a smaller integer specifying the number of encoding functions. The values of feature \(j\) are then encoded using one of the following two methods.
In the first method, \(k_j - 1\) sample points are selected to form \(k_j\) intervals such that \(x_j^{\min} < x_j^{(1)} < x_j^{(2)} < \cdots < x_j^{(k_j-1)} < x_j^{\max}\). The encoding functions are defined with \(x_j^{(0)} = -\infty\), \(x_j^{(k_j)} = \infty\) as follows:
\begin{equation}
\chi_{j}^{s}(x_j) = \begin{cases}
1 \quad & \mathrm{if}\quad x_j^{(s-1)} \leq x_j < x_j^{(s)} \\
0 & \mathrm{otherwise}
\end{cases}
  \label{eq:Appendix4}
\end{equation}
This method models the main effect \(f_{j}(x_j)\) as a step function (piecewise constant function), constant within each of the \(k_j\) intervals defined by consecutive sample points.
In the second method, \(k_j\) sample points are selected such that \(x_j^{\min} = x_j^{(1)} < x_j^{(2)} < \cdots < x_j^{(k_j)} = x_j^{\max}\), and the encoding functions are defined with \(x_j^{(0)} = -\infty\), \(x_j^{(k_j+1)} = \infty\) as follows:
\begin{equation}
\chi_{j}^{s}(x_j) = \begin{cases}
(x_j - x_j^{(s-1)})\;/\;(x_j^{(s)} - x_j^{(s-1)}) \quad & \mathrm{if}\quad {x_j^{(s-1)} \leq x_j < x_j^{(s)}}\\
(x_j^{(s+1)} - x_j)\;/\;(x_j^{(s+1)} - x_j^{(s)}) & \mathrm{if}\quad {x_j^{(s)} \leq x_j < x_j^{(s+1)}}\\
0 & \mathrm{otherwise}
\end{cases}
  \label{eq:Appendix5}
\end{equation}
where \((x_j - x_j^{(0)})\;/\;(x_j^{(1)} - x_j^{(0)})\) and \((x_j^{(k_j+1)} - x_j)\;/\;(x_j^{(k_j+1)} - x_j^{(k_j)})\) are taken to be \(1\). This method models the main effect \(f_{j}(x_j)\) as a piecewise linear function with the sample points serving as the knots, providing constant extrapolation outside the observed range.

Subsequently, let \(\boldsymbol{\beta}\) denote the \(u\)-dimensional vector of all parameters for the main effects:
\begin{equation}
\boldsymbol{\beta} = {\left( \beta_1^1, \beta_1^2, \dots, \beta_1^{k_1}, \beta_2^1, \dots, \beta_d^{k_d} \right)}^\mathrm{T}
  \label{eq:Appendix6}
\end{equation}
where \(u = \sum_{j=1}^{d} k_j\) is the total number of parameters for the main effects.

Let \(X\) be an \(n \times u\) design matrix, whose element in the \(i\)-th row and the column corresponding to parameter \(\beta_j^s\) is \(X_{i,(j,s)} = \chi_j^{s}{(x_{ij})}\), i.e.,
\begin{equation}
X = \begin{pmatrix}
\chi_{1}^1(x_{11}) & \chi_{1}^2(x_{11}) & \dots & \chi_{1}^{k_1}(x_{11}) & \chi_{2}^1(x_{12}) & \dots & \chi_{d}^{k_d}(x_{1d}) \\
\vdots & \vdots & \ddots & \vdots & \vdots & \ddots & \vdots \\
\chi_{1}^1(x_{n1}) & \chi_{1}^2(x_{n1}) & \dots & \chi_{1}^{k_1}(x_{n1}) & \chi_{2}^1(x_{n2}) & \dots & \chi_{d}^{k_d}(x_{nd})
\end{pmatrix}
  \label{eq:Appendix7}
\end{equation}
The centering constraints \(\mathbb{E}\left[f_j(X_j)\right] = 0\) are empirically implemented as \(\sum_{i=1}^n f_j(x_{ij}) = 0\) for each feature \(j\). This translates to the linear constraint \(\sum_{s=1}^{k_j}\beta_j^s(\sum_{i=1}^n\chi_j^s(x_{ij}))=0\) for the parameters \(\{\beta_j^s\}_{s=1}^{k_j}\) associated with feature \(j\). These \(d\) centering constraints can be collectively expressed in matrix form as \(M\boldsymbol{\beta} = \mathbf{0}\). Here, \(M\) is a \(d \times u\) matrix structured such that its \(j\)-th row applies the centering constraint specifically to the parameters of feature \(j\), while setting coefficients for parameters of other features to zero. The overall structure of \(M\) is as follows:
\begin{equation}
M = \begin{pmatrix}
\sum_i\chi_{1}^1(x_{i1}) & \sum_i\chi_{1}^2(x_{i1}) & \cdots & \sum_i\chi_{1}^{k_1}(x_{i1}) & 0 & \cdots & 0 \\
0 & 0 & \cdots & 0 & \sum_i\chi_{2}^1(x_{i2}) & \cdots & 0 \\
\vdots & \vdots & \ddots & \vdots & \vdots & \ddots & \vdots \\
0 & 0 & \cdots & 0 & 0 & \cdots & \sum_i\chi_{d}^{k_d}(x_{id})
\end{pmatrix}
  \label{eq:Appendix8}
\end{equation}
More formally, an element of \(M\) in the \(j\)-th row and the column corresponding to parameter \(\beta_\ell^s\), denoted as \(M_{j,(\ell,s)}\) is defined as
\begin{equation}
M_{j,(\ell,s)} = \begin{cases}
\sum_{i=1}^n \chi_{\ell}^s {(x_{i\ell})} \quad & \mathrm{if} \quad \ell = j \\
0 & \mathrm{if} \quad \ell \neq j
\end{cases}
  \label{eq:Appendix8el}
\end{equation}
Let \(\Delta\) be a \(u \times u\) diagonal matrix for weighted norm minimization, whose diagonal element corresponding to the parameter \(\beta_j^{s}\) is \(\sum_{i=1}^n \chi_j^s(x_{ij})\), i.e.,
\begin{equation}
\Delta = \mathrm{diag} \left( \sum^{n}_{i=1}\chi_{1}^1(x_{i1}),\ \sum^{n}_{i=1}\chi_{1}^2(x_{i1}),\ \cdots,\ \sum^{n}_{i=1}\chi_{d}^{k_d}(x_{id}) \right)
  \label{eq:Appendix9}
\end{equation}
The parameters \(\boldsymbol{\beta}^*\) are found by minimizing \({\left\| {\Delta}^{1/2}{\boldsymbol{\beta}} \right\| }^2\) among solutions to:
\begin{equation}
\mathrm{minimize}\quad {\left\| \tilde{\mathbf{y}} - X {\boldsymbol{\beta}} \right\|}^2\quad \mathrm{subject\;to}\quad M{\boldsymbol{\beta}} = \mathbf{0}
  \label{eq:Appendix10}
\end{equation}

To solve this, let \({\boldsymbol{\gamma}} = {\Delta}^{1/2} \boldsymbol{\beta}\). The problem becomes finding the minimum-norm least squares solution \(\boldsymbol{\gamma}^*\) for the following minimization problem:
\begin{equation}
\mathrm{minimize}\quad {\| \tilde{\mathbf{y}} - (X{\Delta}^{-1/2}) {\boldsymbol{\gamma}} \|}^2\quad \mathrm{subject\;to}\quad \tilde{M}{\boldsymbol{\gamma}} = \mathbf{0}
  \label{eq:Appendix11}
\end{equation}
where \(\tilde{M} = M{\Delta}^{-1/2}\). Then \(\boldsymbol{\beta}^*={\Delta}^{-1/2}\boldsymbol{\gamma}^*\).

If \(\boldsymbol{\gamma} = Z\boldsymbol{\eta}\) where columns of \(Z\) form an orthonormal basis for the null space of \(\tilde{M}\), the problem reduces to an unconstrained minimum-norm least squares problem for \(\boldsymbol{\eta}\):
\begin{equation}
\mathrm{minimize}\quad {\| \tilde{\mathbf{y}} - (X{\Delta}^{-1/2}Z){\boldsymbol{\eta}} \| }^2
  \label{eq:Appendix12}
\end{equation}
The \(u \times (u-\mathrm{rank}(\tilde{M}))\) matrix \(Z\) is found from the SVD of \(\tilde{M} = U \Sigma V^\mathrm{T}\) using the last \(u - \mathrm{rank}(\tilde{M})\) columns of \(V\). The minimum-norm solution \(\boldsymbol{\eta}^*\) is \(\tilde{X}^+\tilde{\mathbf{y}}\) where \(\tilde{X} = X\Delta^{-1/2}Z\) and \(\tilde{X}^+\) is its Moore-Penrose pseudoinverse, typically found via SVD of \(\tilde{X}\).

An alternative for solving the constrained problem, potentially more efficient if forming \(X\Delta^{-1/2}\) is computationally costly, is to solve the following problem:
\begin{equation}
\mathrm{minimize}\quad { \left\| \begin{pmatrix} \tilde{\mathbf{y}} \\ \mathbf{0} \end{pmatrix} - \begin{pmatrix} X\Delta^{-1/2} \\ \kappa \tilde{M} \end{pmatrix} {\boldsymbol{\gamma}}\ \right\|}^2
  \label{eq:Appendix13}
\end{equation}
with \(\kappa > 1\) as a penalty factor for the centering constraints to enforce \(\tilde{M}\boldsymbol{\gamma} \approx \mathbf{0}\).

\hypertarget{second-order-mid}{%
\subsection*{Second-order MID}\label{second-order-mid}}
\addcontentsline{toc}{subsection}{Second-order MID}

The second-order interaction effect \(f_{pq}(x_p,x_q)\) between features \(p\) and \(q\) is modeled using \(k_{p}k_{q}\) parameters \(\left\{\beta_{pq}^{st}\right\}\) and corresponding products of encoding functions as follows:
\begin{equation}
f_{pq}(x_{p},x_{q})=\sum_{s=1}^{k_{p}}\sum_{t=1}^{k_{q}}{\beta_{pq}^{st}}\ \chi_{p}^{s}(x_{p})\ \chi_{q}^{t}(x_{q})
  \label{eq:Appendix14}
\end{equation}
The total number of parameters for all second-order interactions is \(v = \sum_{p}\sum_{q > p}k_{p} k_{q}\), and the collection of all such parameters forms the second-order parameter vector \(\boldsymbol{\beta}'\). The \(n \times v\) design matrix for the second-order interaction effects \(X'\) is constructed such that each column of \(X'\) corresponds to a particular parameter \(\beta^{st}_{pq}\) and consists of the values \(\chi^s_p(x_{ip})\chi^t_q(x_{iq})\) for \(n\) observations. The strict centering constraints are \(\mathbb{E}\left[ f_{pq}(X_p,X_q) \middle| X_p = x_p\right] = 0\) and \(\mathbb{E}\left[ f_{pq}(X_p,X_q) \middle| X_q = x_q\right] = 0\). Empirically, this gives \(k_{p} + k_{q}\) constraints for each feature pair \((p, q)\). The constraint matrix \(M'_{pq}\) for a single interaction \(f_{pq}\) is a \((k_p+k_q) \times k_pk_q\) matrix, whose \(a\)-th row for \(a = 1,\dots,k_p\) enforces \(\sum_{t=1}^{k_q}\beta_{pq}^{at}(\sum_{i}\chi_{p}^a(x_{ip})\chi_q^t(x_{iq})) = 0\) and \((k_p+b)\)-th row for \(b = 1,\dots,k_q\) enforces \(\sum_{s=1}^{k_p}\beta_{pq}^{sb}(\sum_{i}\chi_{p}^s(x_{ip})\chi_{q}^b(x_{iq})) = 0\), i.e.,
\begin{equation}
M'_{pq} = \begin{pmatrix}
\sum_i\chi_{p}^1\chi_{q}^1 & \sum_i\chi_{p}^1\chi_{q}^2 & \cdots & \sum_i\chi_{p}^1\chi_{q}^{k_{q}} & 0 & \cdots & 0 \\
0 & 0 & \cdots & 0 & \sum_i\chi_{p}^2\chi_{q}^1 & \cdots & 0 \\
\vdots & \vdots & \ddots & \vdots & \vdots & \ddots & \vdots \\
0 & 0 & \cdots & 0 & 0 & \cdots & \sum_i\chi_{p}^{k_{p}}\chi_{q}^{k_{q}} \\
\sum_i\chi_{p}^1\chi_{q}^1 & 0 & \cdots & 0 & \sum_i\chi_{p}^2\chi_{q}^1 & \cdots & 0 \\
0 & \sum_i\chi_{p}^1\chi_{q}^2 & \cdots & 0 & 0 & \cdots & 0 \\
\vdots & \vdots & \ddots & \vdots & \vdots & \ddots & \vdots \\
0 & 0 & \cdots & \sum_i\chi_{p}^1\chi_{q}^{k_{q}} & 0 & \cdots & \sum_i\chi_{p}^{k_{p}}\chi_{q}^{k_{q}}
\end{pmatrix}
  \label{eq:Appendix15}
\end{equation}
In the matrix above, \(\sum_i \chi_{p}^{s} \chi_{q}^{t}\) is a shorthand notation for \(\sum_{i=1}^n \chi_{p}^{s}(x_{ip}) \chi_{q}^{t}(x_{iq})\). The full second-order constraint matrix \(M'\) is block diagonal with blocks \(M'_{pq}\) as shown in \eqref{eq:Appendix16}:
\begin{equation}
M' = \begin{pmatrix}
M'_{1,2} & O & \cdots & O \\
O & M'_{1,3} & \cdots & O \\
\vdots & \vdots & \ddots & \vdots \\
O & O & \cdots & M'_{d-1, d}
\end{pmatrix}
  \label{eq:Appendix16}
\end{equation}
Then, the minimization problem for estimating main effects and second-order interactions simultaneously is:
\begin{equation}
\mathrm{minimize}\quad
{ \left\|\ \tilde{\mathbf{y}}-(X, X')\begin{pmatrix} \boldsymbol{\beta} \\ \boldsymbol{\beta}' \end{pmatrix} \right\| }^2\quad
\mathrm{subject\ to}\quad
\begin{pmatrix} M & O \\ O & M' \end{pmatrix} \begin{pmatrix} \boldsymbol{\beta} \\ \boldsymbol{\beta}' \end{pmatrix} = \mathbf{0}
  \label{eq:Appendix17}
\end{equation}
This is solved using the same principles as the first-order case.

\address{%
Ryoichi Asashiba\\
Daido Life Insurance Company\\%
Tokyo, Japan\\
\url{https://github.com/ryo-asashi}\\%
\textit{ORCiD: \href{https://orcid.org/0009-0001-9532-7000}{0009-0001-9532-7000}}\\%
\href{mailto:ryoichi.asashiba@gmail.com}{\nolinkurl{ryoichi.asashiba@gmail.com}}%
}

\address{%
Reiji Kozuma\\
Daido Life Insurance Company\\%
Tokyo, Japan\\
\href{mailto:rktkdm@gmail.com}{\nolinkurl{rktkdm@gmail.com}}%
}

\address{%
Hirokazu Iwasawa\\
Graduate School of Accountancy, Waseda University\\%
Tokyo, Japan\\
\href{mailto:iwahiro@bb.mbn.or.jp}{\nolinkurl{iwahiro@bb.mbn.or.jp}}%
}

\end{article}

\end{document}